\documentclass[10pt, prd, aps, amssymb, amsmath, tightenlines, noshowkeys, 
twocolumn]{revtex4}

\usepackage{graphicx}
\usepackage[T1]{fontenc}
\usepackage[latin1]{inputenc}
\usepackage[english]{babel}
\usepackage{amsmath}
\usepackage{amsfonts}
\usepackage{amssymb}
\usepackage{amsthm}
\usepackage{dsfont}
\usepackage{color}
\usepackage{xcolor}
\usepackage{float}
\usepackage{afterpage}
\usepackage{slashed}
\usepackage[labelformat=parens]{subfig}
\usepackage{bbm}
\usepackage[colorlinks=true, linkcolor=blue, urlcolor=blue, citecolor=blue,
            anchorcolor=blue]{hyperref}
\usepackage[justification=raggedright, font=footnotesize]{caption}

\def\cP{\mathcal P}
\def\cT{\mathcal T}
\def\cPT{\mathcal{PT}}
\def\cH{\mathcal{H}}

\date{\today}

\begin{document}

\title{Non-Hermitian extension of the Nambu--Jona-Lasinio model in 3+1 and 1+1 dimensions}

\author{Alexander Felski}\email{felski@thphys.uni-heidelberg.de}
\author{Alireza Beygi}\email{beygi@thphys.uni-heidelberg.de}
\author{S. P. Klevansky}\email{spk@physik.uni-heidelberg.de}
\affiliation{Institut f\"{u}r Theoretische Physik, Universit\"{a}t Heidelberg,
Philosophenweg 12, 69120 Heidelberg, Germany}

\begin{abstract}

This paper presents a non-Hermitian $\cPT$-symmetric extension of the 
Nambu--Jona-Lasinio (NJL) model of quantum chromodynamics in 3+1 and 1+1 dimensions. 
In 3+1 dimensions, the SU(2)-symmetric NJL Hamiltonian  
$\mathcal{H}_{\textrm{NJL}} = \bar\psi (-i \gamma^k \partial_k + m_0) \psi - 
G [ (\bar\psi \psi)^2 + (\bar\psi i \gamma_5 \vec{\tau} \psi)^2 ]$ is extended
by the non-Hermitian, $\cPT$- and chiral-symmetric bilinear term 
$ig\bar\psi \gamma_5  B_{\mu} \gamma^{\mu} \psi$; 
in 1+1 dimensions, where $\mathcal{H}_{\textrm{NJL}}$ is a form of the 
Gross-Neveu model, it is extended by the non-Hermitian $\cPT$-symmetric but
chiral symmetry breaking term $g \bar\psi \gamma_5 \psi$. 
In each case, the gap equation is derived and the effects of the non-Hermitian
terms on the generated mass are studied. 
We have several findings: in previous calculations for the free Dirac equation
modified to include non-Hermitian bilinear terms, contrary to expectation, no real mass spectrum can be obtained in the chiral limit; 
in these cases a nonzero bare fermion mass is essential for the realization of 
$\cPT$ symmetry in the unbroken regime. 
Here, in the NJL model, in which four-point interactions are present, we 
{\it do} find real values for the mass spectrum also in the limit of vanishing 
bare masses in both 3+1 and 1+1 dimensions, at least for certain specific
values of the non-Hermitian couplings $g$. 
Thus, the four-point interaction overrides the effects leading to $\cPT$ 
symmetry-breaking for these parameter values. 
Further, we find that in both cases, in 3+1 and in 1+1 dimensions, the 
inclusion of a non-Hermitian bilinear term can contribute to the generated 
mass. 
In both models, this contribution can be tuned to be small; 
we thus fix the fermion mass to its  value when $m_0=0$ in the absence of the 
non-Hermitian term, and then determine the value of the coupling required so as to 
generate a bare fermion mass. 
Finally, we find that in both cases, a rich phase structure emerges from the 
gap equation as a function of the coupling strengths.

\end{abstract}

\maketitle

\section{Introduction}
\label{s1}

The study of $\cPT$ symmetry in quantum mechanics has brought to light that 
the combined conditions of invariance under both parity reflection and time
reversal, $ \mathbf{x}\to - \mathbf{x}$ and $t\to -t$, can lead to a real energy spectrum  
\cite{BB}, a fact which today has led to the discovery of many novel and interesting physical 
effects, see, for example references cited in \cite{CMB}. 
The concepts that have been developed have also been extended to non-Hermitian 
bosonic field-theoretic systems, which appear to behave similarly \cite{bhkss}. 
All such systems share the feature that time reversal is even. 
However, non-Hermitian fermionic systems that have {\it odd} time-reversal 
symmetry have much more subtle structures \cite{JSM, AMS, ASR, AS1, AS2, 
BKB}. 

In Ref.~\cite{BKB} we have focused on identifying Lorentz-invariant two-body 
interactions that are $\cPT$ symmetric, but not Hermitian in both 3+1 and 1+1 
 dimensions, and we have investigated the resulting spectra in the context 
of free Dirac-like equations. 
Interestingly, we found there that $\cPT$ symmetry is always realized in 
the broken phase unless we introduce a finite value of the bare or current fermion mass 
$m_0$. 
That is to say, the energy spectrum is always complex, except when the mass 
parameter $m_0\ne0$ exceeds specific model parameter values. 

This leads us to the subject of this current paper: What role do higher-order 
type interactions in fermionic systems play? 
To this end, we study an interacting, relativistic fermionic theory, 
that is extended by including non-Hermitian terms into the Hamiltonian. 
The  Nambu--Jona-Lasinio (NJL) model \cite{NJL}, which has been developed into 
an effective field theory of quantum chromodynamics (QCD), lends itself to 
this. 
One can study how the mechanism of chiral symmetry breaking functions within a 
theory of interacting fermions, and include effects of temperature, density, and 
strong fields \cite{SPK}. 
Its Hamiltonian density, which contains four-point interactions, reads
\begin{equation}
\label{s1e1}
\mathcal{H}_{\textrm{NJL}} = \bar\psi (-i \gamma^k \partial_k + m_0) \psi 
- G [ (\bar\psi \psi)^2 + (\bar\psi i \gamma_5 \vec{\tau} \psi)^2 ],
\end{equation}
where $\gamma$ denotes the Dirac matrices, $\vec{\tau}$ represents the isospin 
SU(2) matrices, and $G$ is a coupling strength. 
The two interaction terms, $(\bar\psi \psi)^2$ and 
$(\bar\psi i \gamma_5 \vec{\tau} \psi)^2$, in the given combination are 
necessary in order to preserve the chiral symmetry of the interaction for the 
two-flavor version of the model. 
On the other hand,  the current quark mass, $m_0$, breaks the chiral 
symmetry of the Hamiltonian explicitly. 

Spontaneous breaking of chiral symmetry occurs via a mechanism that  parallels 
 pairing in the Bardeen-Cooper-Schrieffer (BCS) theory of 
superconductivity \cite{bcs}. 
In the BCS theory, pairing takes place between like particles, that is, 
electrons with opposite spins \cite{leggett}. 
In the NJL model, the pairing takes place between particles and their 
antiparticles, that is, between fermions and antifermions \cite{ASR}. 

The fermions in the 3+1-dimensional NJL model have odd time-reversal symmetry: 
$\cT^2 = -\mathbbm{1}$.  In this work, we extend the NJL model to incorporate non-Hermitian $\cPT$-
symmetric bilinear fermionic terms that also preserve chiral symmetry, and we 
investigate how these terms influence the mass generation. 
We show that, contrary to the results obtained in \cite{BKB}, the extended 
model does admit a real solution for the mass in the chiral limit, when 
$m_0=0$, and that the non-Hermitian  $\cPT$-symmetric bilinear fermionic term 
can be tuned to generate a finite {\it effective current} quark mass, eliminating the 
need for the parameter $m_0$.

We contrast our results from the 3+1-dimensional model 
with those obtained by extending the 1+1-dimensional 
Gross-Neveu model by a non-Hermitian $\cPT$-symmetric bilinear fermionic 
interaction. 
However, in this case, chiral symmetry is broken explicitly by the 
non-Hermitian term, so that the generation of mass is not surprising. 
In addition, this theory behaves as a bosonic theory, since 
$\cT^2=\mathbbm{1}$.

This paper is organized as follows.
In Sec.~\ref{s2}, we study the 3+1-dimensional theory, first recapping the 
symmetry arguments in the context of the equations of motion for the modified
free theory, then solving the gap 
equation and discussing the results. 
The 1+1-dimensional theory is developed in the same fashion in Sec.~\ref{s3}. 
We conclude and summarize our main findings in the final section, Sec.~\ref{s4}.

\section{Non-Hermitian extension of the 3+1-dimensional NJL model}
\label{s2}

We build up the extension of the non-Hermitian NJL model in two stages: First 
we discuss the symmetries associated with including the non-Hermitian chirally 
symmetric bilinear into the free Dirac equation, since this is relevant for 
calculating the associated Green function. Then we set up the new gap equation, 
solve it and present our results. 

\subsection{Symmetries of the free theory modified by an axial-vector bilinear fermionic non-Hermitian term}
\label{s2a}

In \cite{BKB} it was demonstrated explicitly that the bilinear terms 
$i\bar\psi \gamma_5 B_{\mu} \gamma^{\mu} \psi$ and 
$i\bar \psi T_{\mu\nu}\sigma^{\mu\nu}\psi$ are both non-Hermitian and 
invariant under the combination of parity reflection and time reversal. 
In this section, we will consider only the former term, 
$\Gamma=ig\bar\psi  \gamma_5 B_{\mu} \gamma^{\mu} \psi$, 
since (as we shall show) it is also invariant under a chiral transformation. 
Notationally, we use the Dirac representation of the gamma matrices \cite{BD},
$$\gamma^0 = \left( \begin{array}{cc} \mathbbm{1} & 0\\ 0 & -\mathbbm{1} 
\end{array} \right), \quad \quad \gamma^k = \left( \begin{array}{cc} 0 & 
\sigma^k\\ -\sigma^k & 0 \end{array} \right),$$ 
where $\sigma^k$ are the Pauli matrices and 
$\gamma_5 = i \gamma^0 \gamma^1 \gamma^2 \gamma^3$.

Combining the free Dirac Hamiltonian 
$\cH_F = \bar\psi(-i \gamma^k\partial_k +m_0) \psi $ with the bilinear 
interaction $\Gamma$ results in the modified free Hamiltonian
\begin{equation}
\label{s2e1}
\cH_0 = \cH_F+\Gamma =\bar\psi (-i \gamma^k \partial_k + m_0 
+ ig  \gamma_5 B_{\mu}\gamma^{\mu}) \psi,
\end{equation}
from which one infers the equation of motion 
\begin{equation}
\label{s2e2}
(i \slashed\partial - m_0 - ig  \gamma_5 B_{\mu}\gamma^{\mu}) 
\psi(t, \mathbf{x}) = 0.
\end{equation}

The usual parity and time-reversal operations for the spinors are defined as
\cite{BD},
\begin{equation}
\label{s2e3}
\cP: \psi(t, \mathbf{x}) \to \cP \psi(t, \mathbf{x}) \cP^{-1} = 
\gamma^0 \psi(t, -\mathbf{x}),
\end{equation}
\begin{equation}
\label{s2e4}
\cT: \psi(t, \mathbf{x}) \to \cT \psi(t, \mathbf{x}) \cT^{-1} = 
Z \psi^*(-t, \mathbf{x}),
\end{equation}
where $Z = i \gamma^1 \gamma^3$.  
We note that (\ref{s2e4}) implies $\cT^2 = -\mathbbm{1}$, that is, the time-reversal 
operator in 3+1 dimensions is odd.

By setting $\mathbf{x}\to -\mathbf{x}$ in (\ref{s2e2}), we have 
\begin{equation}
\label{s2e5}
\big(i \gamma^0 \partial_0 - i \gamma^k \partial_k
-m_0 - ig \gamma_5 B_0\gamma^0 - ig \gamma_5 B_k \gamma^k \big)
\psi(t, -\mathbf{x}) = 0.
\end{equation}
Multiplying (\ref{s2e5}) from the left by $\gamma^0$, we obtain
\begin{equation}
\label{s2e6}
[i \slashed\partial - m_0 + ig( \gamma_5 B_0 \gamma^0 - \gamma_5 B_k \gamma^k )]
\gamma^0\psi(t, -\mathbf{x}) = 0,
\end{equation}
where we have used the fact that $\gamma^0$ anti-commutes with 
$\gamma^k $
and $\gamma_5$. 
Equation (\ref{s2e6}) implies that (\ref{s2e2}) is not form invariant under parity 
reflection. 
On the other hand, taking the complex conjugate of (\ref{s2e6}) and replacing $t\to -t$  
 gives
\begin{equation}
\label{s2e7}
\begin{split}
\big[ &i \gamma^0 \partial_0 -i \gamma^1 \partial_1 
+ i \gamma^2 \partial_2 -i \gamma^3 \partial_3 - m_0  -ig\gamma_5 B_0 \gamma^0 \\
&+ig\gamma_5B_1 \gamma^1 -ig\gamma_5 B_2\gamma^2+ig\gamma_5B_3\gamma^3 \big] 
 \gamma^0\psi^*(-t, \mathbf{-x})
 = 0,
\end{split}
\end{equation}
where we have used $(\gamma^2)^* = - \gamma^2$. 
Multiplying this expression with $i \gamma^1 \gamma^3$ from the left does lead to
 a form-invariant Dirac equation,
\begin{equation}
(i \slashed\partial - m_0 - ig \gamma_5  B_{\mu}\gamma^{\mu}) 
\gamma^0[i\gamma^1\gamma^3]\psi^*(-t, -\mathbf{x}) 
= 0
\end{equation}
or 
\begin{equation}
\label{s2e8}
(i \slashed\partial - m_0 - ig \gamma_5 B_{\mu} \gamma^{\mu}) 
\mathcal{PT} \psi(t, \mathbf{x}) 
= 0.
\end{equation}
The form invariance of (\ref{s2e2}) under the combined space-reflection and time-reversal symmetries implies that the 
spectrum of the modified free Hamiltonian  (\ref{s2e1})  can be real.

Furthermore, in the chiral limit of vanishing bare mass $m_0$ Eq.~(\ref{s2e2}) 
becomes
\begin{equation}
\label{s2e9}
(i \slashed\partial - ig  \gamma_5 B_{\mu}\gamma^{\mu}) \psi (t, \mathbf{x})= 0,
\end{equation}
which respects continuous chiral symmetry: under the transformation, 
$\psi \to e^{i \alpha \gamma_5} \psi$, where $\alpha \in \mathbb{R}$, 
(\ref{s2e9}) becomes
\begin{equation}
\label{s2e10}
(i \slashed\partial - ig  \gamma_5 B_{\mu}\gamma^{\mu}) e^{i \alpha \gamma_5} 
\psi = 0.
\end{equation}
Multiplying from the left by $e^{i \alpha \gamma_5}$, we find
\begin{equation}
\label{s2e11}
(\mathbbm{1} + i \alpha \gamma_5 + ...) (i \slashed\partial 
- ig  \gamma_5 B_{\mu}\gamma^{\mu}) e^{i \alpha \gamma_5} \psi = 0,
\end{equation}
which reduces to
\begin{equation}
\label{s2e12}
(i \slashed\partial - ig  \gamma_5 B_{\mu}\gamma^{\mu}) e^{-i \alpha \gamma_5} 
e^{i \alpha \gamma_5} \psi = 0,
\end{equation}
so that we recover (\ref{s2e9}).

\subsection{The gap equation for the non-Hermitian NJL model}
\label{s2b}

We define our non-Hermitian NJL model to be
\begin{widetext}
\begin{equation}
\label{s2e13}
\cH= \cH_0 - G [ (\bar\psi \psi)^2 + (\bar\psi i \gamma_5 \vec{\tau} \psi)^2 ]
=\bar\psi (-i \gamma^k \partial_k + m_0+ig \gamma_5 B_{\mu}\gamma^{\mu} ) \psi 
- G [ (\bar\psi \psi)^2 + (\bar\psi i \gamma_5 \vec{\tau} \psi)^2 ]  ,
\end{equation}
\end{widetext}
based on the modified free non-Hermitian Hamiltonian $\cH_0$ in (\ref{s2e1}).

Following Feynman-Dyson perturbation theory, the full propagator $S$ can be 
expressed in terms of the free propagator $S^{(0)}$ and the proper self-energy 
$\Sigma^*$ through the (algebraic) Dyson equation as
\begin{equation}
\label{s2e14}
i S_{\alpha \beta}(k) = i S^{(0)}_{\alpha \beta}(k) 
+ [i S^{(0)}_{\alpha \lambda}(k)] [-i \Sigma^*_{\lambda \mu}(k)] 
[i S_{\mu \beta}(k) ],
\end{equation}
where, in this case, $S^{(0)}$ is associated with $\cH_0$ and is given formally as 
\begin{equation}
\label{s2e15}
S^{(0)}(k) = (\slashed k -m_0 -ig\gamma_5 B_\mu \gamma^\mu)^{-1}.
\end{equation}

The approximation to the proper self-energy to first order in an expansion in 
$1/N_c$  (the Hartree approximation) is improved on through imposing a 
self-consistency condition, that is,  the free propagator, $S^{(0)}$, is 
replaced by the full one, $S$. 
With this prescription, the self-consistent proper self-energy takes the form
\begin{equation}
\label{s2e16}
\Sigma^{sc}_{\lambda \mu}(k) =
2 i G\delta_{\lambda \mu} N_c N_f \int \frac{d^4 p}{(2 \pi)^4} \, 
{\rm tr}[S(p)],
\end{equation}
where $N_c$ and $N_f$ are the number of colors and flavors, respectively, tr 
denotes the spinor trace, and $\mu$ and $\lambda$ are spin indices.  
One sees that in this approximation $\Sigma^{sc}(k)$ is a constant, so that one 
may identify 
 \begin{equation}
\label{s2e17}
\Sigma^{sc}_{\lambda \mu}(k) = (m^* - m_0) \delta_{\lambda \mu},
\end{equation}
where $m^*$ plays the role of an effective mass. 
Thus, we obtain the same structure for the gap equation as in the usual NJL 
case \cite{SPK},
\begin{equation}
\label{s2e18}
m^* = m_0 + 2 i G N_c N_f \int \frac{d^4 p}{(2 \pi)^4} \, {\rm tr}[S(p)].
\end{equation}

In the free extended theory, $S^{(0)}$ satisfies
\begin{equation}
\label{s2e19}
(\slashed k- m_0 - ig \gamma_5  B_{\mu}\gamma^{\mu}) 
S^{(0)}_{\alpha \beta}(k) =  \delta_{\alpha \beta}.
\end{equation}
By acting with the same operator on the 
Dyson equation (\ref{s2e14}) one finds
\begin{equation}
\label{s2e20}
(\slashed k - m^*- i g \gamma_5 B_{\mu}\gamma^{\mu}) 
S_{\alpha \beta}(k) =  \delta_{\alpha \beta}.
\end{equation}
This implies that the full propagator which is required to determine the 
solutions of the gap equation, is just the free propagator with the mass shift, 
$m_0 \rightarrow m^*$.
Thus, in order to set up the gap equation, we need to insert the 
propagator of the free non-Hermitian theory (\ref{s2e15}) into (\ref{s2e16}) 
and evaluate the spinor trace.

The method involves recasting $S^{(0)}(p)$ as determined by 
(\ref{s2e15}) in an algebraic form that has a scalar denominator. 
Thus, we first expand $S^{(0)}(p)$ with the factor 
$(\slashed{p} + m_0 + i g \gamma_5 B_\mu \gamma^\mu)$ such that the denominator 
takes the form
\begin{align}
\label{s2e21}
\begin{split}
&(\slashed{p} - m_0 - i g \gamma_5 B_\mu \gamma^\mu) (\slashed{p} + m_0 
+ i g \gamma_5 B_\nu \gamma^\nu)\\
&=
p^2 - m_0^2 - g^2 B \cdot B - 2 i g m_0 \gamma_5 B_\mu \gamma^\mu 
- 2 i g B\cdot p \gamma_5,
\end{split}
\end{align}
where we have used the fact that
$\gamma_5 \gamma^{\mu} \slashed{p} - \slashed{p} \gamma_5 \gamma^{\mu} = 
2 p^{\mu} \gamma_5$
and the Minkowski inner product is denoted with a dot.
The second and the third terms in (\ref{s2e21}) are, however, still not scalar, 
so we expand the result with a new factor containing the opposite signs in 
those two terms. 
This leads to
\begin{align*}
&[ p^2 - m_0^2 - g^2 B \cdot B - 2 i g m_0 \gamma_5 B_\mu \gamma^\mu 
- 2 i g B \cdot p \gamma_5 ]\\
&\times
[ p^2 - m_0^2 - g^2 B \cdot B + 2 i g m_0 \gamma_5 B_\mu \gamma^\mu 
+ 2 i g B \cdot p \gamma_5 ]\\
&=
(p^2 - m_0^2 - g^2 B \cdot B)^2 - 4 g^2 m_0^2 B \cdot B 
+ 4 g^2 (B \cdot p)^2
\end{align*}
for the denominator.
Then the free propagator for the non-Hermitian Hamiltonian takes the form
\begin{align}
\label{s2e22}
\begin{split}
&S^{(0)}(p) = [ \slashed{p} + m_0 + i g \gamma_5 B_\mu \gamma^\mu ]\\
&\times
\frac{[ p^2 - m_0^2 - g^2 B \cdot B + 2 i g m_0 \gamma_5 B_\mu \gamma^\mu 
+ 2 i g B \cdot p \gamma_5  ]}
{ (p^2 - m_0^2 - g^2 B \cdot B)^2 - 4 g^2 m_0^2 B \cdot B
+ 4 g^2 (B \cdot p)^2}.
\end{split}
\end{align}
By performing the trace, most of the matrix terms in the numerator vanish and 
we find 
\begin{widetext}
\begin{equation}
\label{s2e23}
\mathrm{tr}[ S^{(0)}(p) ] = \frac {4 m_0 (p^2 - m_0^2 + g^2 B \cdot B)}
{(p^2 - m_0^2 - g^2 B \cdot B)^2 - 4 g^2 m_0^2 B \cdot B 
+ 4 g^2 (B \cdot p)^2}. 
\end{equation}

As is argued in Eqs.~(\ref{s2e19}) and (\ref{s2e20}) and the discussion 
following, the full propagator $S(p)$ (and its trace) can be obtained from the 
free propagator $S^{(0)}(p)$ by replacing the bare mass $m_0$ by the effective 
mass $m^*$. 
Thus, the gap equation, (\ref{s2e18}), becomes
\begin{equation}
\label{s2e24}
m^* = m_0 + \frac{8 i G N_c N_f m^*}{(2 \pi)^4}  I_4,
\end{equation}
where
\begin{equation}
\label{s2e25}
I_4 = \int d^4 p \,\frac{  p^2 - {m^*}^2 + g^2 B \cdot B}{
 (p^2 - {m^*}^2 - g^2 B \cdot B)^2 - 4 g^2 {m^*}^2 B \cdot B
+ 4 g^2 (B \cdot p)^2}.
\end{equation}
\end{widetext}

At this point it is necessary to specify a regularization scheme in order to 
evaluate the momentum integral, $I_4$. 
Noting that the general results for the (standard) NJL model are qualitatively 
insensitive to the scheme used, we choose the Euclidean four-momentum cutoff 
method. 
We thus transform to Euclidean coordinates and introduce a radial four-momentum 
Euclidean cutoff $\Lambda$. 
That is, $p_0 = i p_4$ and $B_0 = i B_4$ such that 
$p_E^2 = p_1^2+ ... + p_4^2 = -p^2$ and $B_E^2 = -B \cdot B$. 
In the spherical coordinate system with zenith in the direction along $B_E$, 
the Euclidean product
\begin{equation}
\label{s2e26}
B_E \cdot p_E = \lvert B_E \rvert \lvert p_E \rvert \cos\theta,
\end{equation}
contains only the zenithal angle $\theta$. 
After introducing the radial cutoff $\Lambda$, the momentum integral becomes
\begin{align}
\label{s2e27}
\begin{split}
I_4 &= - i \int_0^\Lambda dr \int_0^\pi d\theta \int_0^\pi d\varphi_1 
\int_0^{2 \pi} d\varphi_2 \, r^3 \sin^2\theta \sin\varphi_1\\
\times& \frac{( r^2 + {m^*}^2 + g^2 B_E^2 )}{
 (r^2 + {m^*}^2 - g^2 B_E^2)^2 + 4 g^2 {m^*}^2 B_E^2 + 4 g^2 B_E^2 r^2 
\cos^2\theta }
.
\end{split}
\end{align}
The $\varphi_1$ and $\varphi_2$ integrations are readily evaluated and we find 
\begin{equation}
\label{s2e28}
I_4 = -4 i \pi \!\int_0^\Lambda \!\! dr r^3 \frac{r^2 + {m^*}^2 + g^2 B_E^2}
{4 g^2 B_E^2 r^2} \int_0^\pi \!\!
d\theta \frac{\sin^2\theta}{A(r) - \sin^2\theta},
\end{equation}
where $A(r) = (r^2 + {m^*}^2 + g^2 B_E^2)^2 / 4g^2 B_E^2 r^2$.
Using 
\begin{equation}
\label{s2e29}
\int_0^\pi d\theta \, \frac{\sin^2\theta}{A(r) - \sin^2\theta} = \pi 
\Bigg( \sqrt{\frac{A(r)}{A(r) - 1}} - 1 \Bigg),
\end{equation}
for the angular integral, we find 
\begin{align}
\label{s2e30}
\begin{split}
I_4 = 4 i \pi^2 &\int_0^\Lambda dr \, \frac{r (r^2 + {m^*}^2 + g^2 B_E^2)}
{4 g^2 B_E^2}\\
&\times
\bigg( 1 - \frac{r^2 + {m^*}^2 + g^2 B_E^2}{\sqrt{(r^2 + {m^*}^2 
+ g^2 B_E^2)^2 - 4 g^2 B_E^2 r^2}} \bigg).
\end{split}
\end{align}
The radial integration can now be performed (see Ref.~\cite{GR}), leading to
\begin{widetext}
\begin{align}
\label{s2e31}
\begin{split}
I_4 = -\frac{i \pi^2}{4 g^2 B_E^2} \bigg\{
&-\Lambda^4 + \Lambda^2 \Bigl( \sqrt{(\Lambda^2 + {m^*}^2 - g^2 B_E^2)^2 
+ 4g^2 B_E^2 {m^*}^2} - 2({m^*}^2 + g^2 B_E^2) \Bigr)\\
&
+({m^*}^2 + 7 g^2 B_E^2) \Bigl( \sqrt{(\Lambda^2 + {m^*}^2 - g^2 B_E^2)^2 
+ 4 g^2 B_E^2 {m^*}^2} -({m^*}^2 + g^2 B_E^2) \Bigr)\\
&
+ 4 g^2 B_E^2 (2 g^2 B_E^2 - {m^*}^2) \ln\Big[ \frac{1}{2 {m^*}^2} 
\Big( \sqrt{(\Lambda^2 + {m^*}^2 - g^2 B_E^2)^2 + 4 g^2 B_E^2 {m^*}^2} 
+ \Lambda^2 + {m^*}^2 - g^2 B_E^2 \Big) \Big] \bigg\}.
\end{split}
\end{align}

In the limit of vanishing bare mass $m_0$, and introducing the dimensionless 
scaled quantities $\tilde{m} = m^* / \Lambda$, 
$\tilde{g} = g \lvert B_E \rvert / \Lambda$, and $\tilde{G} = G \Lambda^2$, the 
gap equation (\ref{s2e24}) for the non-Hermitian extension of the NJL model can be recast in the form
\begin{align}
\label{s2e32}
\begin{split}
\frac{2 \pi^2}{\tilde{G} N_c N_f} = \frac{1}{4 \tilde{g}^2}
\bigg\{&
\sqrt{(1 + \tilde{m}^2 - \tilde{g}^2)^2 +4 \tilde{g}^2 \tilde{m}^2} 
(1 + \tilde{m}^2 + 7 \tilde{g}^2) - (\tilde{m}^2 + \tilde{g}^2)(2 + \tilde{m}^2 
+ 7 \tilde{g}^2) - 1\\
&
+4 \tilde{g}^2 (2 \tilde{g}^2 - \tilde{m}^2) \ln\Big[ \frac{1}{2 \tilde{m}^2} 
\Big( \sqrt{(1 + \tilde{m}^2 - \tilde{g}^2)^2 + 4 \tilde{g}^2 \tilde{m}^2} 
+ 1 + \tilde{m}^2 - \tilde{g}^2 \Big) \Big] \bigg\}.
\end{split}
\end{align}
\end{widetext}

\begin{figure}
\centering
\subfloat[$0 \leq \tilde{g} \leq 0.1$]{
\includegraphics[width=0.45\textwidth]
{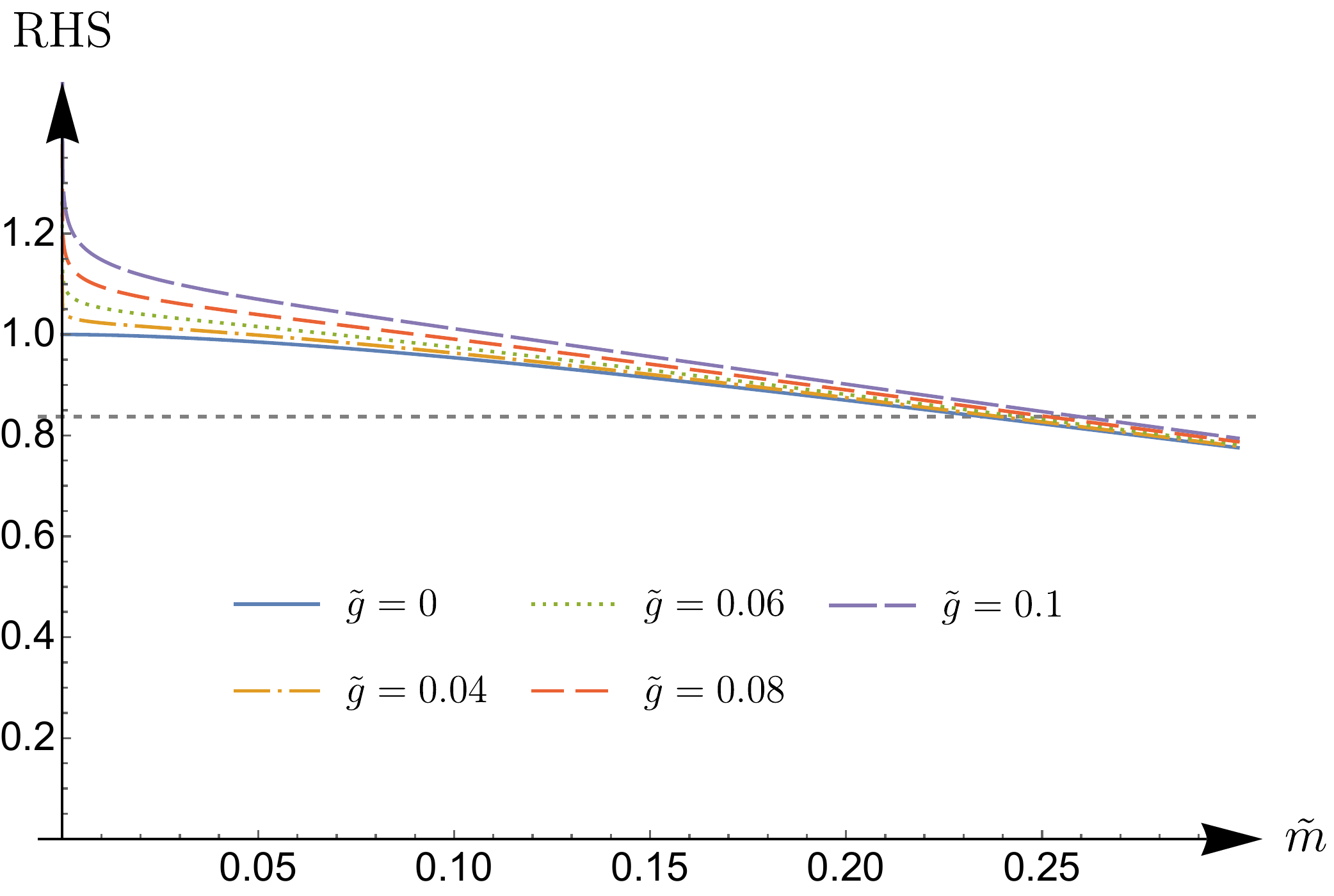}
\label{f1a}
}
\hfill
\subfloat[$0.1 \leq \tilde{g} \leq 0.8$]{
\includegraphics[width=0.45\textwidth]
{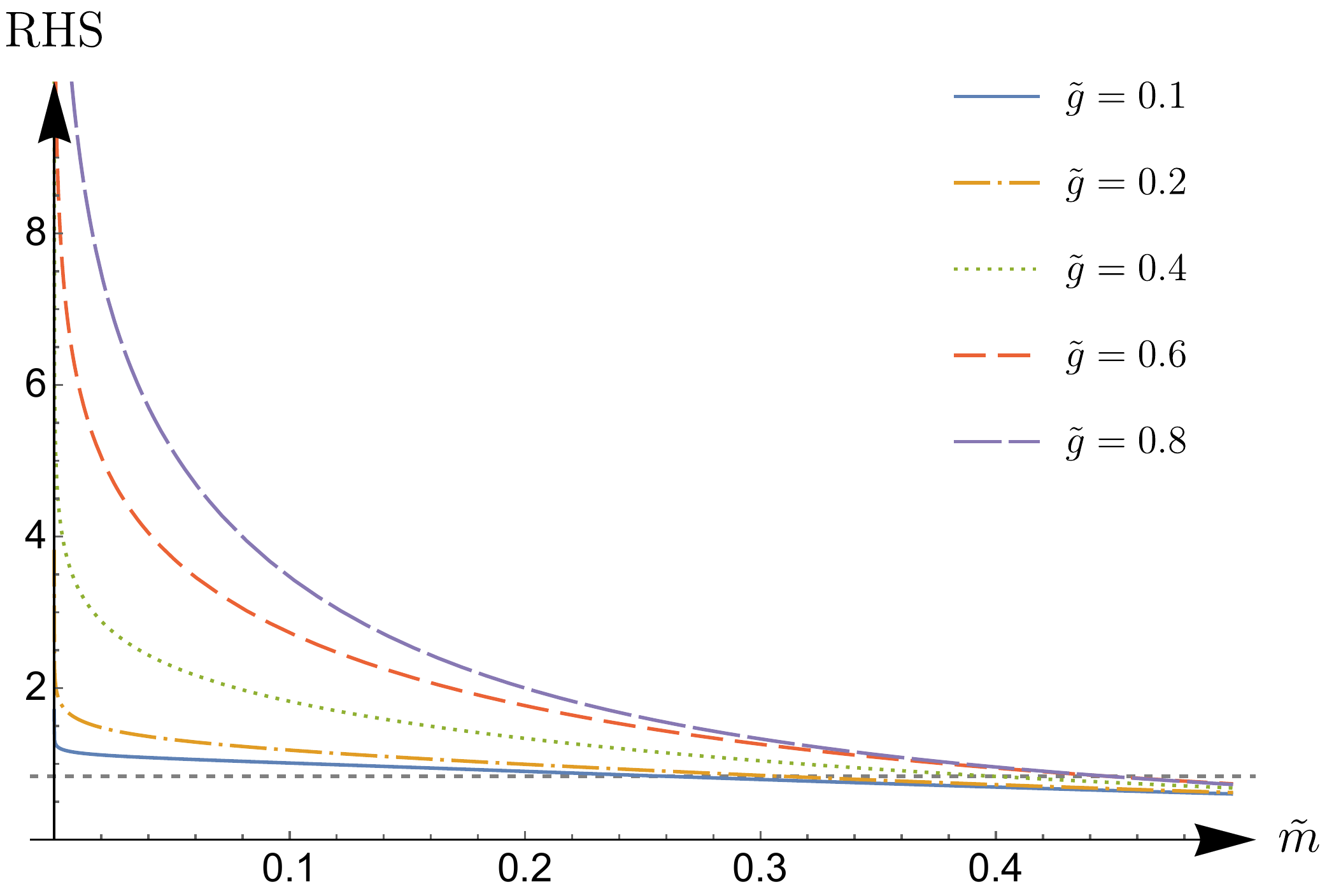}
\label{f1b}
}
\hfill
\subfloat[$1 \leq \tilde{g} \leq 1.6$]{
\includegraphics[width=0.45\textwidth]
{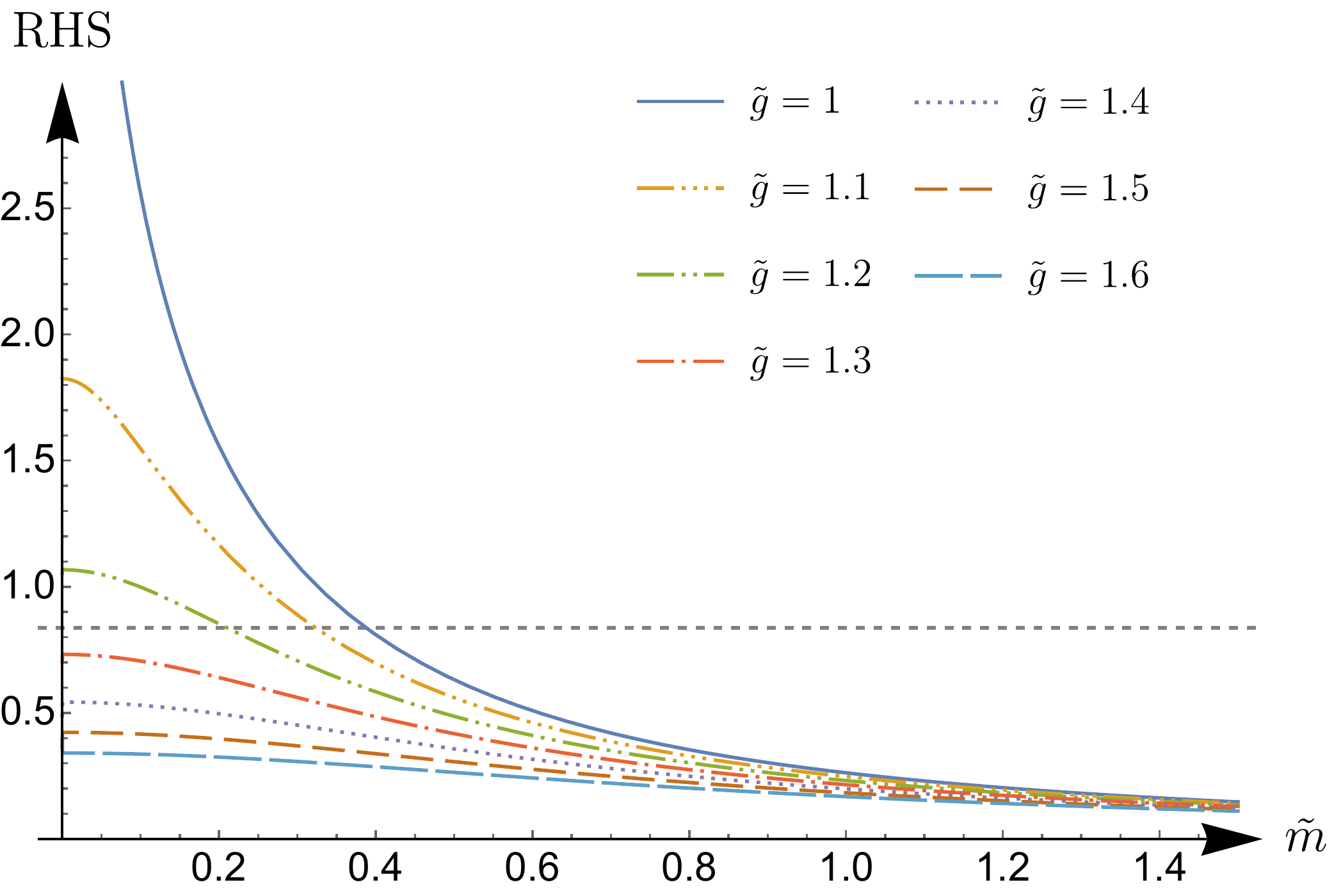}
\label{f1c}
}
\caption{
\label{f1}
Behavior of the right hand side of (\ref{s2e32}) as a function of the scaled 
mass $\tilde{m}$ for given ranges of the scaled coupling constant $\tilde{g}$ 
(curves). 
The constant ${2 \pi^2}/{\tilde{G} N_c N_f} $ is plotted as a dotted line for 
fixed values of $G$ and $\Lambda$.
}
\end{figure}
This is a central result.  
We note that in the limit $\tilde{g} \to 0$, that is, the limit in which the 
non-Hermitian term vanishes, we recover the known gap equation of the 
conventional Hermitian NJL model in this regularization scheme \cite{SPK},
\begin{equation}
\label{s2e33}
\frac{2 \pi^2}{G \Lambda^2 N_c N_f} = 1 - \frac{{m^*}^2}{\Lambda^2} 
\ln\bigg( 1 + \frac{\Lambda^2}{{m^*}^2} \bigg).
\end{equation}

In analyzing the new gap equation for the non-Hermitian NJL Hamiltonian, we choose 
the cutoff and the four-point interaction strength to be $\Lambda = 1015$ MeV 
and $G \Lambda^2 = 3.93$, taking these values from the Hermitian conventional 
model for which $\tilde g=0$. 
We determine the solutions of the gap equation (\ref{s2e32}) from the 
intersection of the function given by the right hand side of the equation with 
the (real positive) constant on the left hand side.

\begin{figure*}
\centering
\null\hfill
\subfloat[]{
\includegraphics[width=0.45\textwidth]
{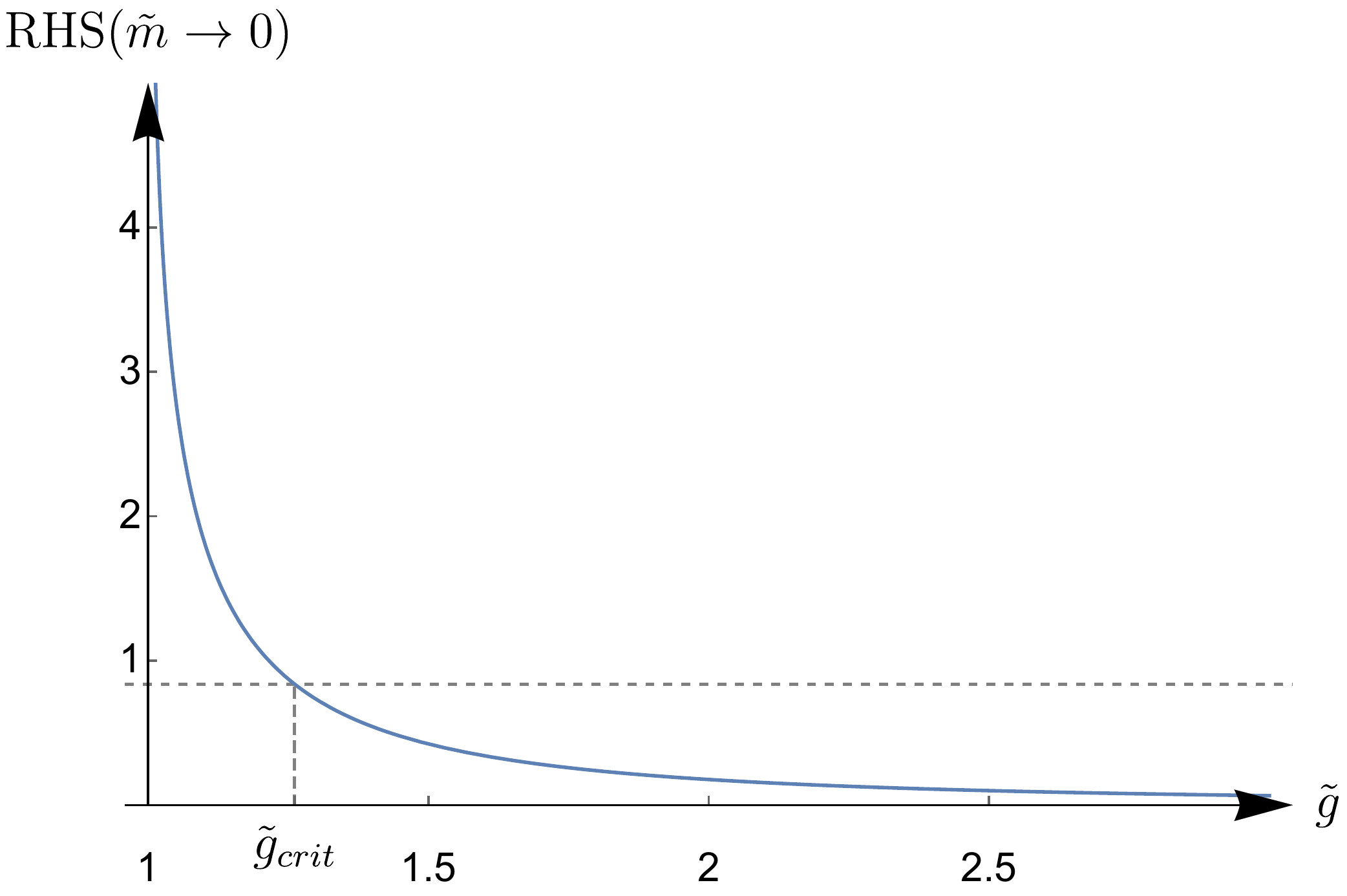}
\label{f2a}
}
\hfill
\subfloat[]{
\includegraphics[width=0.45\textwidth]
{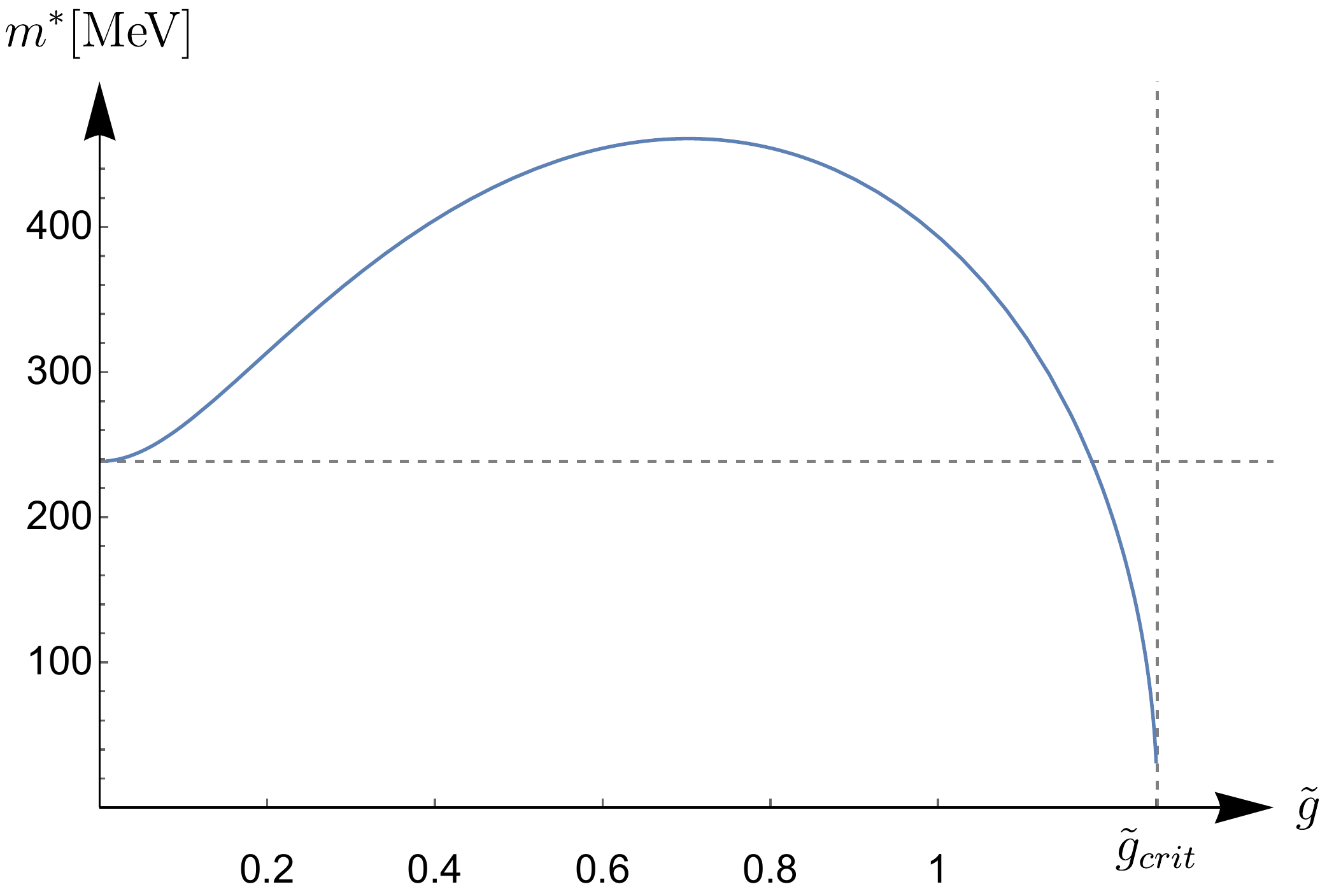}
\label{f2b}
}
\hfill\null
\caption{ 
\label{f2}
The behavior of the right hand side of (\ref{s2e32}) in the limit 
$\tilde{m}\to 0$ as a function of $\tilde{g}$ is shown in \ref{f2a}. 
In \ref{f2b} the mass solution to the gap equation is shown as a function of 
the scaled coupling constant $\tilde{g}$.
}
\end{figure*}

Figure \ref{f1} shows the behavior of the right hand side of (\ref{s2e32}) as a 
function of $\tilde{m}$ for different ranges of $\tilde{g}$. 
This function evaluates to purely real positive values and vanishes 
for large values of $\tilde m$. 
In Fig.~\ref{f1a}, when $\tilde{g}=0$ the right hand side of (\ref{s2e32}) takes on a finite value at $\tilde{m}=0$ and leads to the standard real solution of the conventional NJL theory. 
In the range $0 <\tilde{g} \leq 1$ there is always a singularity at $\tilde{m} = 0$, so that the gap equation  has a real solution in this region,
see in particular Figs.~\ref{f1a} and \ref{f1b}. 
However, for large values of the coupling strength $\tilde{g} > 1$ the 
function has only a finite maximum at vanishing mass, see Fig.~\ref{f1c}.  
We determine the height of this maximum as a function of 
$\tilde{g}$, see Fig.~\ref{f2a}, and find that for coupling values 
$\tilde{g} > \tilde g_{\textrm {crit}} \approx 1.261$ it lies below the value of the 
constant given by the left hand side of (\ref{s2e32}), $2\pi^2/\tilde G N_cN_f$. 
Therefore, no real solution to the gap equation can be found in this region.
For coupling values $1< \tilde g < \tilde g_{\textrm {crit}}$, a real solution 
does, however, exist and can be found again as the intersection of the right 
hand side of (\ref{s2e32}) with the constant on the left hand side of this 
equation. 

The real mass solution to (\ref{s2e32}) (given in MeV) is shown in 
Fig.~\ref{f2b} as a function of the scaled coupling constant $\tilde{g}$.
The solution starts at $m^* \approx 238.487$ MeV when $\tilde g=0$ and rises to a maximum 
value of $m^* \approx 460.870$ MeV when $\tilde g\approx 0.702$.  
Thereafter it falls to zero at $\tilde g_{\textrm {crit}}$ so that the solution 
for a particular value of the mass is doubly degenerate. 

In particular, we can determine the range of scaled coupling values $\tilde g$ 
that would be necessary to generate an effective bare quark mass 
that corresponds to the values of the actual bare up or down quark masses.
A mass solution to the gap equation that would generate a mass 
difference in the range of the bare up quark mass, that is $m_u=(1.7-3.3)$ MeV, 
compared to the Hermitian theory (with $\tilde g=0$),
requires a rescaled coupling constant in the range $\tilde{g}\approx(0.025-0.034)$ or $\tilde{g}\approx(1.181-1.182)$.
For the range of the bare down quark mass $m_d=(4.1-5.8)$ MeV we obtain 
the coupling range $\tilde{g}\approx(0.038-0.046)$ or $\tilde{g}\approx(1.179-1.181)$.
We expect the coupling to be small, and therefore lie in the first range given. 

We thus conclude that the inclusion of a non-Hermitian $\cPT$-symmetric, chirally invariant 
term into the NJL Hamiltonian serves to increase the mass that is 
generated dynamically. 
That is, it can, for a certain parameter range, account for the extra small 
average value of $\sim 5$ MeV that is usually specified as a parameter 
for the bare quark mass.

\section{Non-Hermitian extension of the 1+1-dimensional Gross-Neveu model}
\label{s3}

The standard Gross-Neveu model is essentially the NJL model 
(\ref{s1e1}) defined in 1+1 dimensions, taken without isospin 
$\vec{\tau}$, and originally also excluding the second four-point axial 
interaction term.
The latter version has a discrete chiral symmetry \cite{GN}, while the former 
has this symmetry promoted to a continuous chiral symmetry. 
Here we can consider (\ref{s1e1}) as it stands with 
$\vec{\tau}$ also in 1+1 dimensions, as isospin plays a vital role in 
phenomenological applications. 
But this is simply cosmetic, as the second interaction term plays no role in 
the derivation of the gap equation to leading order in the $1/N_c$ expansion 
and has the same form for all these models.

Our intention is again to introduce a bilinear non-Hermitian term into the 
Gross-Neveu model and contrast the results obtained with those found in 3+1 
dimensions. 
In this case, the only non-Hermitian $\cPT$-symmetric  bilinear available is  
$\Gamma = g \bar\psi \gamma_5 \psi$. 
As in Sec.~\ref{s2}  we first recall the symmetry properties associated with 
this term within the modified free theory, before analyzing the gap equation for the 
fully interacting system.

\subsection{Symmetries of the free theory modified by a pseudoscalar bilinear fermionic non-Hermitian terrm}
\label{s3a}

In 1+1 dimensions, the only bilinear term that is non-Hermitian and $\mathcal{PT}$ symmetric is 
the pseudoscalar $\Gamma = g \bar\psi \gamma_5 \psi$.  
Using the representation for the Dirac matrices 
\cite{AAR},
$$\gamma^0 = \left( \begin{array}{cc} 0 & 1\\ 1 & 0 \end{array} \right), 
\quad \quad \gamma^1 = \left( \begin{array}{cc} 0 & 1\\ -1 & 0 \end{array} 
\right),$$ 
where $(\gamma^0)^2 = \mathbbm{1}$, $(\gamma^1)^2 = -\mathbbm{1}$, and 
$\gamma_5 = \gamma^0 \gamma^1$,
we consider the modified free Hamiltonian 
\begin{equation}
\label{s3e1}
\mathcal{H}_0 = \cH_F + \Gamma=\bar \psi(-i\gamma^1\partial_1 + m_0 + g\gamma_5) \psi,
\end{equation}
in which 
$\cH_F=\bar\psi (-i \gamma^1\partial_1 + m_0) \psi$.  
In addition to being non-Hermitian, $\Gamma$ also breaks the individual symmetries of parity reflection and 
time reversal; it is, however, invariant under the combined operations, namely, 
$\cPT$ symmetry. 
This can be seen by studying the symmetries of the
equation of motion associated with (\ref{s3e1}),
\begin{equation}
\label{s3e2}
(i \slashed\partial - m_0 - g \gamma_5) \psi(t, x) = 0.
\end{equation}
Under a parity transformation, the spinor transforms as \cite{BKB},
\begin{equation}
\cP: \psi(t, x) \to \cP \psi(t, x) \cP^{-1} = \gamma^0 \psi(t, -x),
\end{equation}
and under time reversal \cite{BKB} as
\begin{equation}
\label{s3e5}
\cT: \psi(t, x) \to \cT \psi(t, x) \cT^{-1} = \gamma^0 \psi^*(-t, x),
\end{equation}
which implies that $\cT^2 = +\mathbbm{1}$ for Dirac fermions in 1+1 dimensions.

Setting $x\to -x$ in (\ref{s3e2}) leads to
\begin{equation}
\label{s3e3}
\big( i \gamma^0 \partial_0 - i \gamma^1 \partial_1 - m_0 - g \gamma_5 \big)
 \psi(t, -x) = 0,
\end{equation}
where $\partial_0 = \partial_t$ and $\partial_1 = \partial_x$ and multiplying (\ref{s3e3}) 
from the left by $\gamma^0$, we obtain
\begin{equation}
\label{s3e4}
(i \slashed\partial - m_0 + g \gamma_5) \gamma^0\psi(t, -x) = 0,
\end{equation}
where we have used the fact that $\gamma^0$ anti-commutes with $\gamma^1$ and 
$\gamma_5$.
From Eq.~(\ref{s3e4}), one sees that the last term is odd under a parity 
transformation. Now letting $t\to-t$ and taking the complex conjugate of 
 (\ref{s3e4}), we have
\begin{equation}
\label{s3e6}
\big[ i \gamma^0 \partial_0 -i \gamma^1 \partial_1 - m_0 + g \gamma_5 
\big] \gamma^0 \psi^*(-t, -x) =0.
\end{equation}
By multiplying (\ref{s3e6}) from the left by $\gamma^0$, we establish 
form invariance with the original equation (\ref{s3e2}),
\begin{equation}
\label{s3e7}
(i \slashed\partial - m_0 - g \gamma_5) \gamma^0\gamma^0 \psi^*(-t, -x) 
= 0.
\end{equation}
We recognize $ \gamma^0\gamma^0 \psi^*(-t, -x) = \mathcal{PT} \psi(t, x)$.
As a result, although the equation of motion is not separately invariant under 
parity reflection and time reversal, it remains invariant under the combined 
operations of $\cP$ and $\cT$. 
This fact suggests once again that the modified free non-Hermitian Hamiltonian (\ref{s3e1}) can 
have a real spectrum \cite{BJR}. 
If we iterate (\ref{s3e2}), we obtain the two-dimensional Klein-Gordon equation 
as
\begin{equation}
\label{s3e8}
(\partial^2 + m_0^2 - g^2) \psi = 0,
\end{equation}
where we have used the fact that $\slashed\partial^2 = \partial^2$. 
Equation (\ref{s3e8}) implies that the propagated mass is shifted by $g^2$, and 
is real and nonzero only if $m_0^2 > g^2$.

The non-Hermitian $\cPT$-symmetric mass term suggested here also breaks the 
discrete and continuous chiral symmetry explicitly, in apposition to the case 
in 3+1 dimensions: 
in the limit of vanishing bare mass $m_0$ (\ref{s3e2}) reads 
\begin{equation}
\label{s3e9}
(i \slashed\partial - g \gamma_5) \psi = 0.
\end{equation}
Under the discrete chiral transformation $\psi \to \gamma_5 \psi$, this becomes
\begin{equation}
\big[ i \gamma^0 \partial_0 + i \gamma^1 \partial_1 - g \gamma_5 \big] \gamma_5 
\psi = 0,
\end{equation}
which by multiplying by $\gamma_5$ from the left, turns into
\begin{equation}
\label{s3e10}
(i \slashed\partial + g \gamma_5) \psi = 0,
\end{equation}
where we have used the facts that $\{ \gamma_5, \gamma^{\mu} \} = 0$ and 
$(\gamma_5)^2 = \mathbbm{1}$. 
Equation (\ref{s3e10}) shows the non-invariance of (\ref{s3e9}) under the 
discrete chiral transformation; similarly it is not invariant under 
a continuous chiral transformation where $\psi \to e^{i \alpha \gamma_5} \psi$ 
for some real $\alpha$.

\subsection{The gap equation for the non-Hermitian Gross-Neveu model}
\label{s3b}

\begin{figure}
\centering
\subfloat[$\tilde{g} = 0.5$]{
\includegraphics[width=0.45\textwidth]
{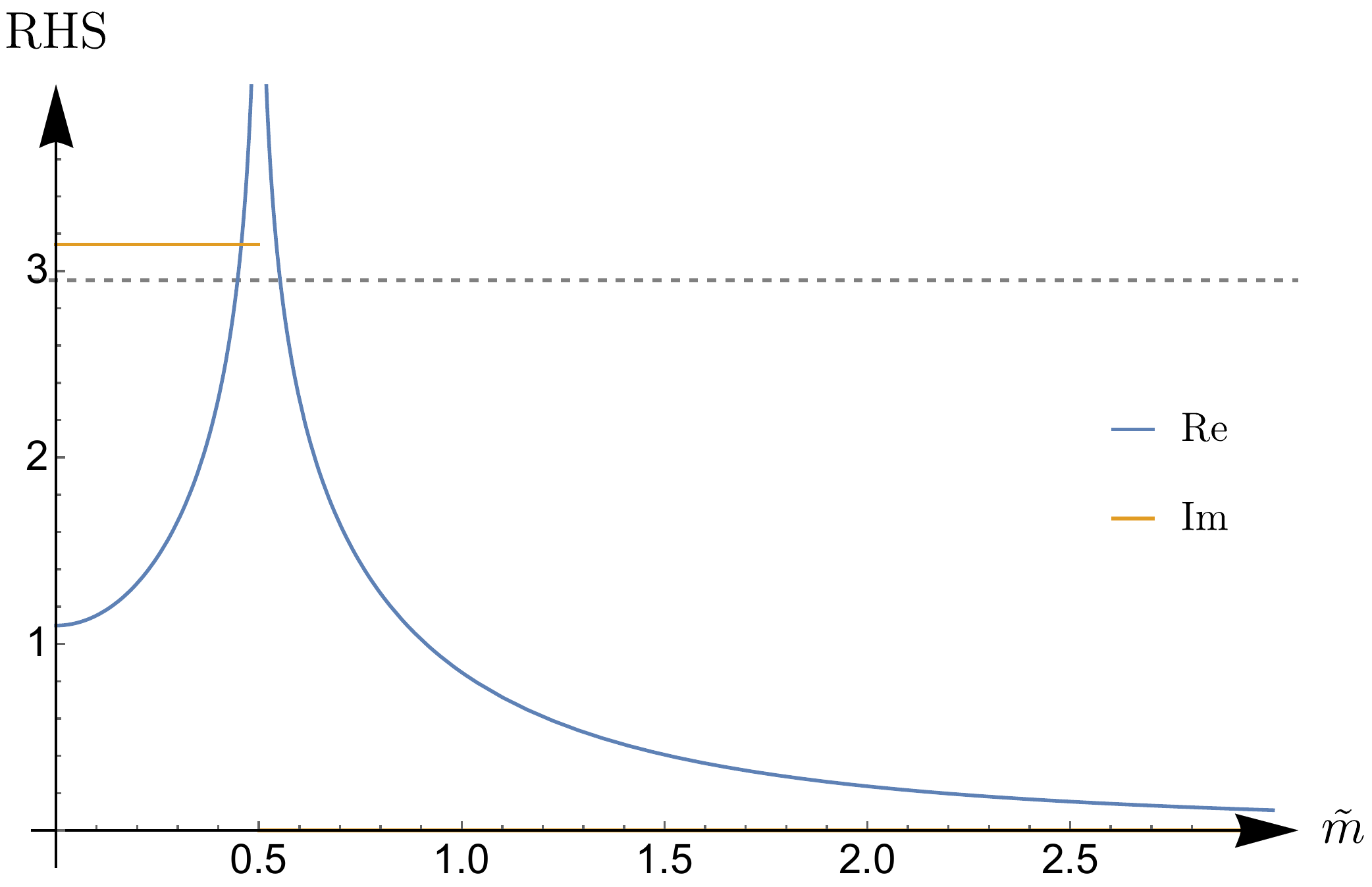}
\label{f3a}
}
\hfill
\subfloat[$\tilde{g} = 1$]{
\includegraphics[width=0.45\textwidth]
{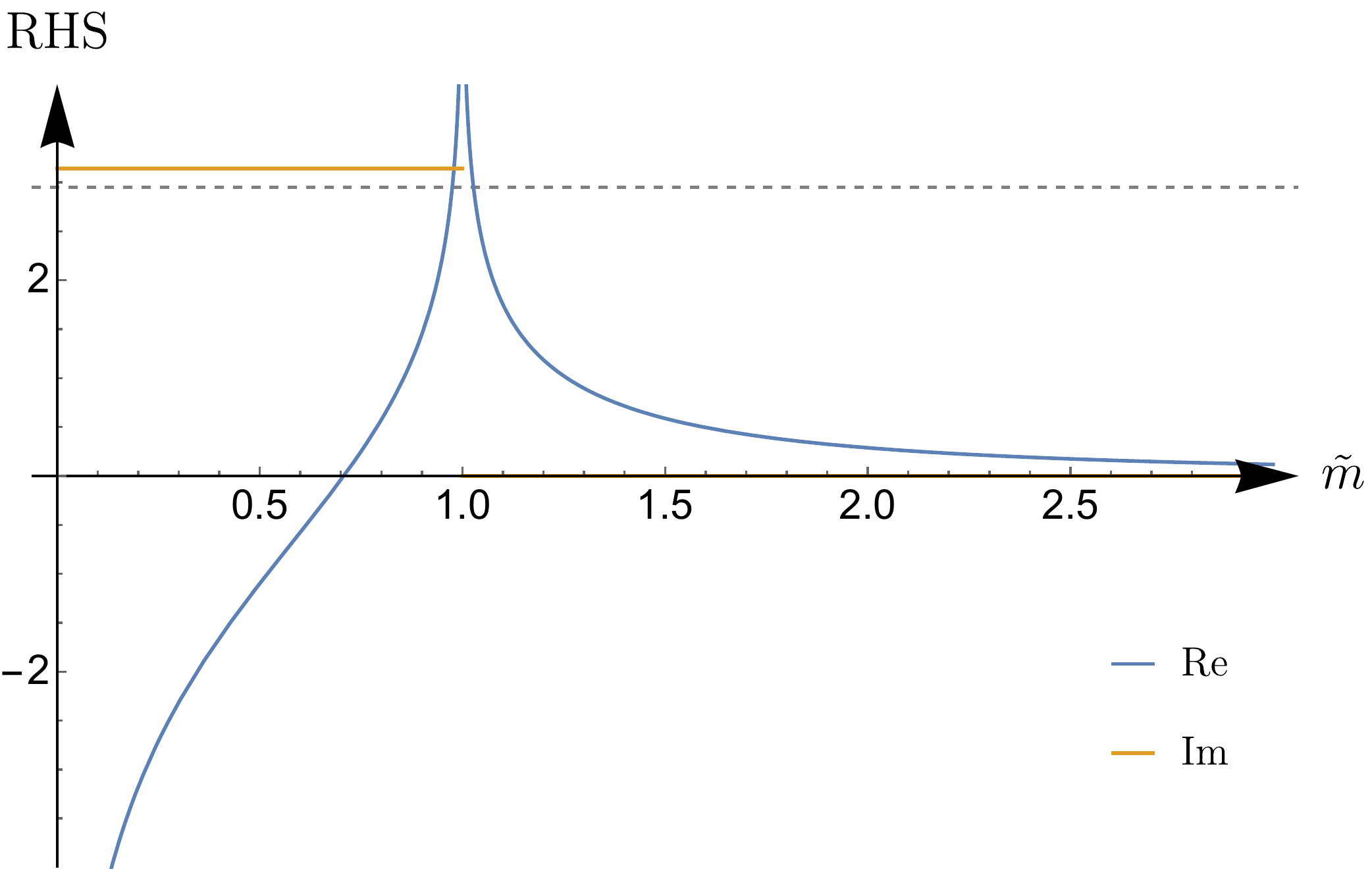}
\label{f3b}
}
\hfill
\subfloat[$\tilde{g} = 2$]{
\includegraphics[width=0.45\textwidth]
{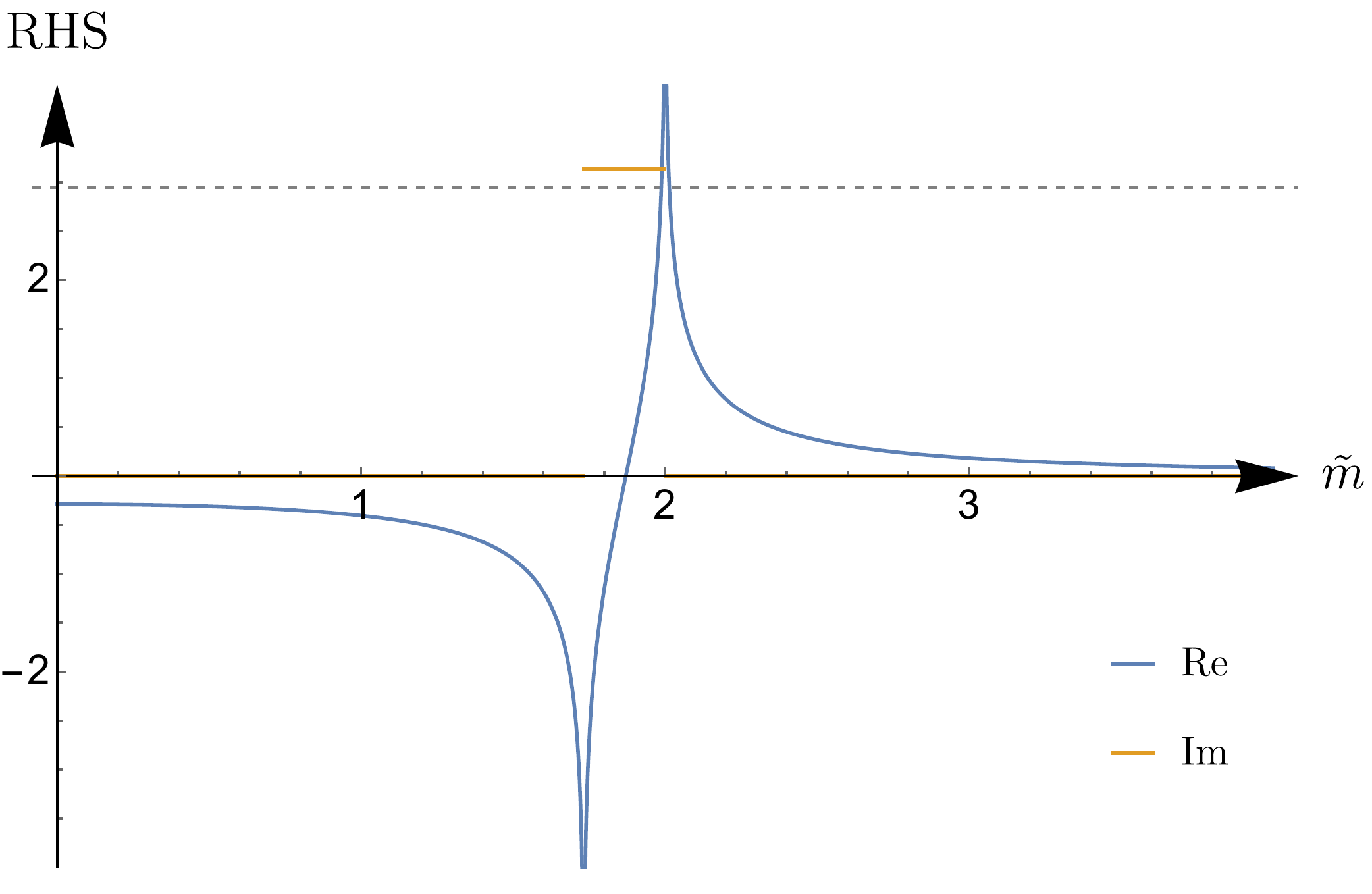}
\label{f3c}
}
\caption{
\label{f3}
Behavior of the right hand side of the gap equation~(\ref{s3e19}) as a function of the scaled mass $\tilde{m}$ for different 
values of the scaled coupling constant $\tilde{g}$.
}
\end{figure}

We define the 
non-Hermitian Gross-Neveu model in 1+1 dimensions as
\begin{widetext}
\begin{equation}
\label{s3e11}
\cH= \cH_0 - G [ (\bar\psi \psi)^2 + (\bar\psi i \gamma_5 \vec{\tau} \psi)^2 ]
= \bar\psi (-i \gamma^1 \partial_1 + m_0 + g \gamma_5) \psi - G [ (\bar\psi 
\psi)^2 + (\bar\psi i \gamma_5 \vec{\tau} \psi)^2 ],
\end{equation}
\end{widetext}
where $\mathcal{H}_0$  is the modified free non-Hermitian Hamiltonian in (\ref{s3e1}).
The associated free propagator 
$S^{(0)}$  is given formally as
\begin{equation}
\label{s3e12}
S^{(0)}(p) = (\slashed{p} - m_0 - g \gamma_5)^{-1} .
\end{equation}
The arguments leading to the gap equation (\ref{s2e18}) in 3+1 dimensions are 
applicable here as well, so that in 1+1 dimensions, the gap equation reads
\begin{equation}
\label{s3e13}
m^* = m_0 + 2 i G N_c N_f \int \frac{d^2 p}{(2 \pi)^2} \, {\rm tr}[S(p)],
\end{equation}
where the full propagator $S(p)$ has the same form as $S^{(0)}$, but with 
$m_0$ replaced by $m^*$. 
We thus proceed to evaluate the trace by expanding (\ref{s3e12})
with $(\slashed{p} + m_0 + g \gamma_5)$ and taking $m_0\to m^*$ so that the 
denominator becomes
\begin{align}
\label{s3e14}
\begin{split}
&(\slashed{p} - m^* - g \gamma_5) (\slashed{p} + m^* + g \gamma_5)\\
&= p^2 - m^*{}^2 - g^2 - 2 m^* g \gamma_5 - 2 g \gamma_5 \slashed{p},
\end{split}
\end{align}
and expanding again with opposite sign in the last two terms then leads to the 
full propagator 
\begin{align}
\label{s3e15}
\begin{split}
&S(p) =\\
&\frac{(\slashed{p} + m^* + g \gamma_5) [ p^2 - m^*{}^2 - g^2 
+ 2 m^* g \gamma_5 + 2 g \gamma_5 \slashed{p} ]}
{(p^2 - m^*{}^2 + g^2)^{2}},
\end{split}
\end{align}
with the trace
\begin{equation}
\label{s3e16}
\mathrm{tr}[S(p)] = \frac{2 m^*{}}{p^2 - m^*{}^2 + g^2}.
\end{equation}
Thus, the gap equation (\ref{s3e13}) becomes
\begin{equation}
\label{s3e17}
m^* = m_0 + \frac{4 i G N_c N_f m^*}{(2 \pi)^2}  I_2,
\end{equation}
where
\begin{equation}
\label{s3e18}
I_2 = \int d^2 p \, \frac{1}{p^2 - {m^*}^2 + g^2}.
\end{equation}
Introducing Euclidean coordinates with $p_0 = i p_2$, transforming to spherical 
coordinates and introducing a radial cutoff $\Lambda$, $I_2$ becomes
\begin{align}
\label{s3e19}
\begin{split}
I_2 &= -i \int_0^{2 \pi} d\theta \int_0^{\Lambda} dr \, \frac{r}{r^2 + {m^*}^2 
- g^2}\\
&= -i \pi \ln\bigg( 1 + \frac{\Lambda^2}{{m^*}^2 - g^2} \bigg).
\end{split}
\end{align}
This leads to the gap equation,  
\begin{equation}
\label{s3e19}
\frac{\pi}{G N_c N_f} = \ln\Big( 1+ \frac{1}{\tilde{m}^2 - \tilde{g}^2} \Big),
\end{equation}
in terms of the scaled mass $\tilde{m} = m^* / \Lambda$ and the scaled coupling 
constant $\tilde{g} = g / \Lambda$,  in the limit of vanishing bare mass 
$m_0$.
Here, we consider the case of two flavors, $N_f = 2$, and three colors, 
$N_c = 3$.  
We note that the four-point interaction strength in the gap equation 
(\ref{s3e19}) does not scale with the cutoff length $\Lambda$. 
This is to be expected in the 1+1-dimensional NJL model.

\begin{figure}
\centering
\null\hfill
\includegraphics[width=0.45\textwidth]
{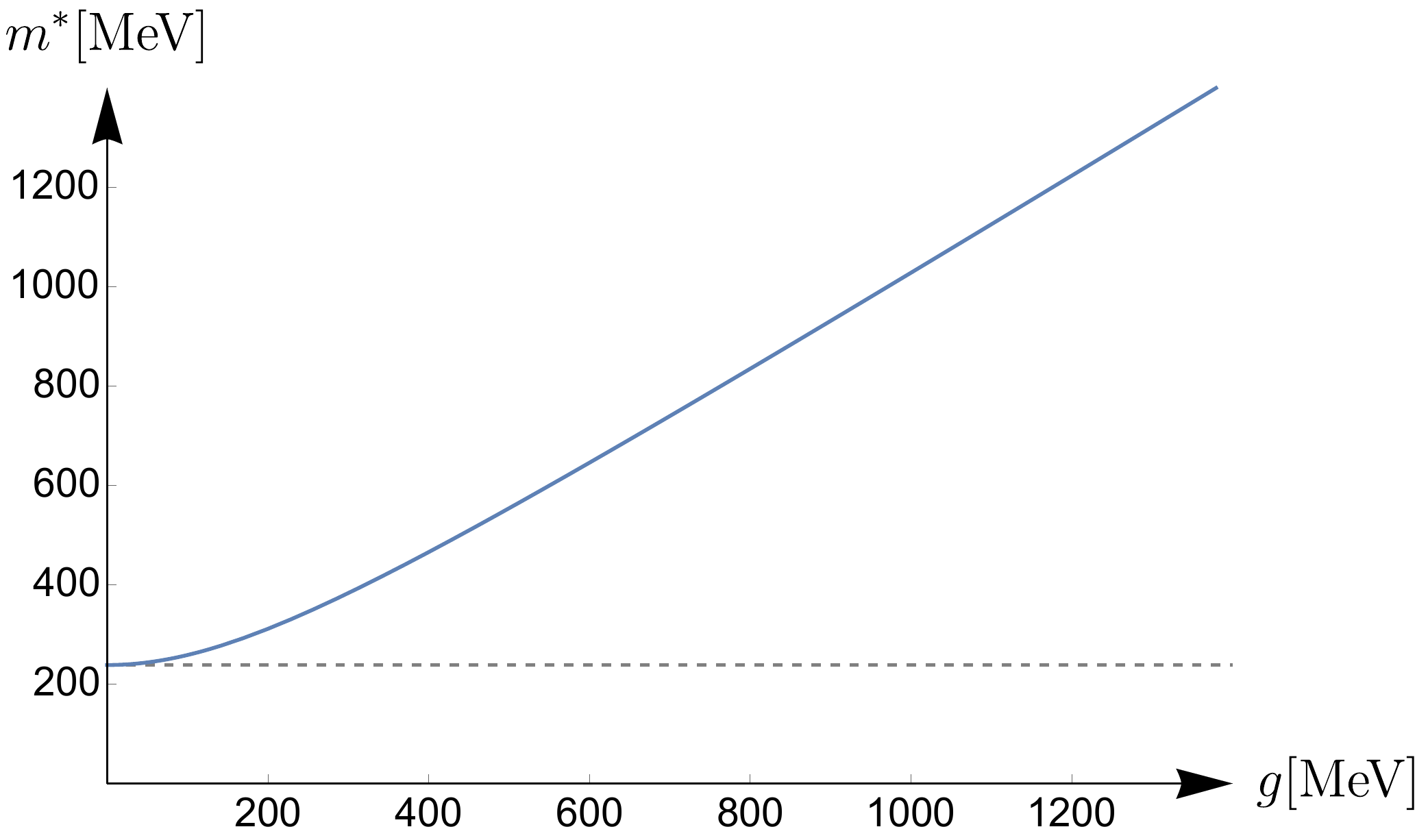}
\hfill\null
\caption{ 
\label{f4}
Visualization of the mass solution $m$ as a 
function of the coupling constant $g$.
}
\end{figure}

In order to obtain scaled mass solutions to the gap equation that are
comparable with those of the 3+1-dimensional model, we fix the 
four-point interaction strength $G$ such that the solution of 
the Hermitian theory (at $g=0$) is identical to the solution obtained there, 
namely $\tilde{m}=0.235 $. 
This is achieved with $G\approx0.177$.

The left hand side of (\ref{s3e19}) is a real positive constant and the 
solutions of the gap equation can be determined as the intersection of the 
function given by the right hand side of (\ref{s3e19}) with this constant. 
In Fig.~\ref{f3}, the behavior of this function with the scaled mass $\tilde{m}$ 
is shown for fixed values of the scaled coupling constant $\tilde{g}$. 
It visualizes the following properties of the function:

(a) The right hand side of (\ref{s3e19}) generally has two singularities, 
namely, one at $\tilde{m}^2 = \tilde{g}^2$ and one at 
$\tilde{m}^2 = \tilde{g}^2 - 1$. 
Inbetween, the function has complex values. 
As a result, the gap equation does not have real solutions in the range 
$\sqrt{\tilde{g}^2 - 1} \leq \tilde{m} \leq \tilde{g}$. 

(b) For the special choice of coupling values $0 < \tilde{g} \leq 1$, the 
second singularity does not occur for real masses and therefore the function 
has complex values for all masses in the range 
$0 \leq \tilde{m} \leq \tilde{g}$. 

(c) For coupling constants $\tilde{g} > 1$ the function takes on real, but 
negative, values for masses lying below the first singularity, 
$0 \leq \tilde{m} \leq \sqrt{\tilde{g}^2 - 1}$, so that an intersection with 
the positive constant given by the left hand side of (\ref{s3e19}) is not 
possible. 

For mass values $\tilde{m}<\tilde g$, only complex solutions exist; $\cPT$ symmetry
is realized in the broken phase.
For mass values $\tilde{m} > \tilde{g}$, however, the function on the right hand side of 
(\ref{s3e19}) generally takes 
on all real positive values larger than zero. 
Therefore, a real mass solution to the gap equation is guaranteed to exist in 
the region $\tilde{m}^2 > \tilde{g}^2$ for all scaled coupling values 
$\tilde{g}$.  $\cPT$ symmetry is manifestly realized.
We compare this with the analysis presented in 
Sec.~\ref{s3a}. 
Therein, we found that real mass solutions exist only if $m_0^2 > g^2$ (see 
Eq.~(\ref{s3e8})), which implies that we are in the region of unbroken $\cPT$ 
symmetry. 
Here, in the limit of vanishing bare mass $m_0$, we obtain a similar 
relation for the existence of the mass solutions of the gap equation.

Figure~\ref{f4} shows the behavior of the solution for  $m$ as function of the  coupling constant $g$.

Analogous to the 3+1-dimensional model we can determine the 
range of the scaled coupling constant $\tilde{g}$ that gives rise to 
mass solutions of the gap equation which describe the range of the 
scaled mass $\tilde{m}$ corresponding to a current up or down quark mass
in the 3+1-dimensional model. 
Namely, the scaled-mass range $\tilde{m}=(0.237-0.238)$, corresponding to 
the up quark mass in the 3+1-dimensional model, is generated by scaled 
coupling values in the range $\tilde{g}\approx(0.028-0.039)$.
The scaled-mass range $\tilde{m}=(0.239-0.240)$, corresponding to the down 
quark mass in the 3+1-dimensional model, is generated by scaled coupling 
values in the range $\tilde{g}\approx(0.044-0.052)$.

\subsection{Renormalization}
\label{s3c}

\begin{figure}
\centering
\null\hfill
\includegraphics[width=0.45\textwidth]
{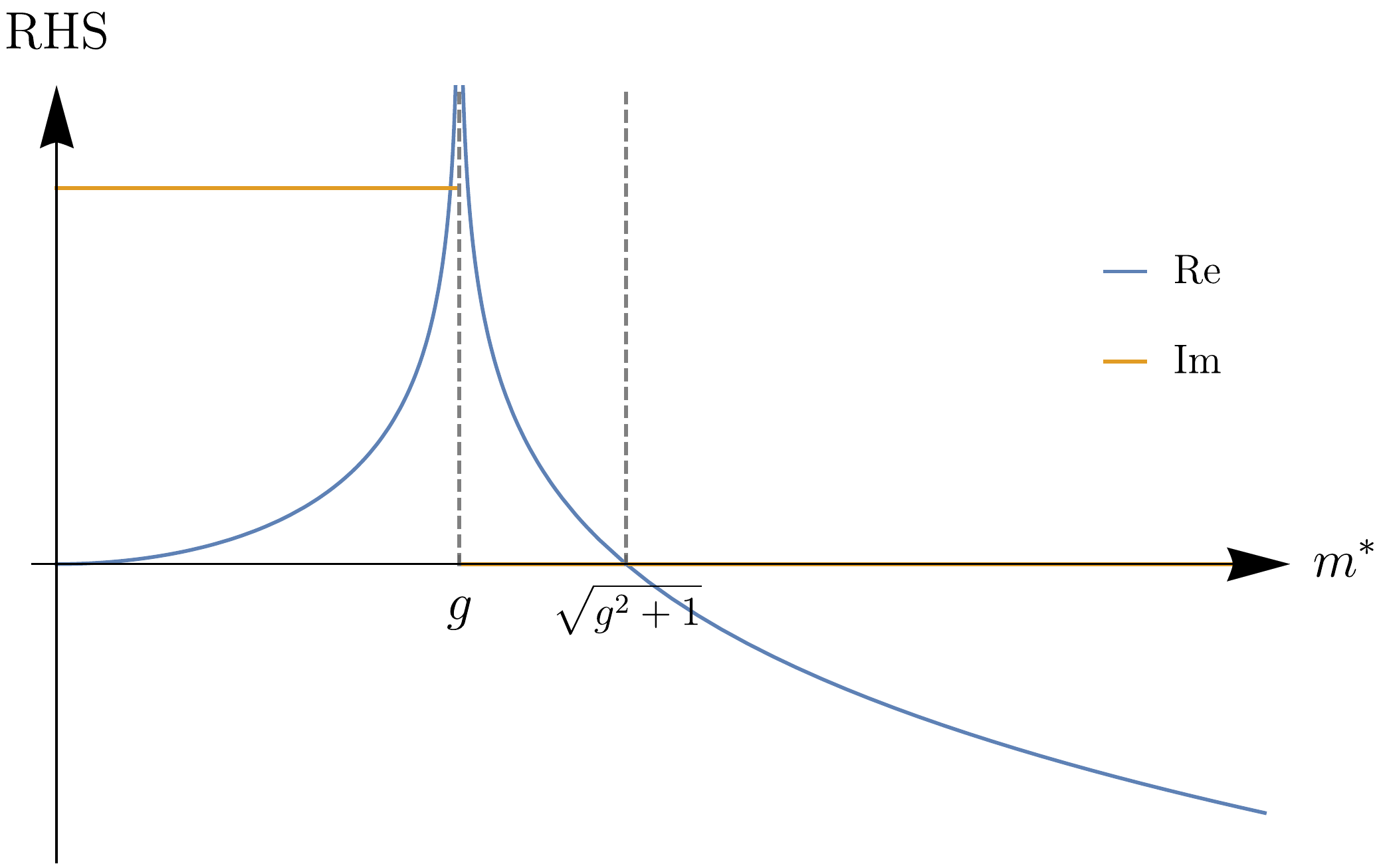}
\hfill\null
\caption{ 
\label{f5}
Behavior of the right hand side of the renormalized gap equation (\ref{s3e21})
as a function of the effective mass $m^*$.
}
\end{figure}

Contrary to the 3+1-dimensional NJL model, the Gross-Neveu model in 
1+1 dimensions is renormalizable \cite{GN}. 
This is already indicated by the fact that the four-point interaction 
strength $G$ is a dimensionless parameter, see Eq.~(\ref{s3e19}). 
Thus we can absorb the ultraviolet divergence occurring in the limit of a large
cutoff $\Lambda$ into the four-point interaction strength $G$ 
to find the gap equation of the renormalized theory.

Introducing the arbitrary dimensionful (energy) scale $c= 1$MeV writing  $\Lambda = c \lambda$, with
$\lambda$ dimensionless, we expand  (\ref{s3e19})
in the limit $\lambda \to \infty$, yielding 
\begin{equation}
\label{s3e20}
\frac{\pi}{G N_c N_f} = 
2\ln\lambda + \ln\Big(\frac{c^2}{{m^*}^2 - g^2}\Big)
+ O\Bigl(\frac{1}{\lambda^2} \Bigr),
\end{equation}
and  absorb the divergent first term on the right hand side into a renormalized  
four-point interaction strength $G_R$ defined as
\begin{equation}
\frac{1}{G_R}= \frac{1}{G} -\frac{2N_c N_f}{\pi} \ln\lambda.
\end{equation}
In the limit $\lambda \to \infty$ keeping $G_R$ fixed, we thus obtain
the renormalized gap equation
\begin{equation}
\label{s3e21}
\frac{\pi}{G_R N_c N_f} = \ln\Big(\frac{c^2}{{m^*}^2-g^2} \Big) .
\end{equation}

Figure \ref{f5} shows the behavior of the right hand side of (\ref{s3e21}) as a function of the effective
mass $m^*$. This function has a singularity at $m^* = g$ and a root
at $m^* = \sqrt{g^2+1}$.
For $m^*<g$ the function takes on complex values and for $m^*>\sqrt{g^2+1}$
it evaluates to negative real values.
In particular, all possible real mass solutions to the renormalized gap equation satisfy the relation $m^*>g$ since the left hand side of (\ref{s3e21}) is a real constant.

It is instructive to calculate the mass solution of (\ref{s3e21}) as a function of $g$,
\begin{equation}
\label{s3e22}
m^* = \sqrt{g^2+c^2\Bigl[\exp\Bigl(\frac{\pi}{G_RN_cN_f}\Bigr)\Bigr]^{-1}},
\end{equation}
and compare it to the scaled mass solution $\tilde{m}=m^*/\Lambda$ obtained from the unrenormalized gap equation 
(\ref{s3e19}),
\begin{equation}
\tilde{m} = \sqrt{\tilde{g}^2+\Bigl[\exp\Bigl(\frac{\pi}{GN_cN_f}\Bigr) -1\Bigr]^{-1}} ,
\end{equation}
in terms of the scaled coupling $\tilde{g}=g/\Lambda$.
In both cases the term in square brackets is a fixed positive constant 
determined by the value of the four-point interaction strength $G_R$ or 
$G$ respectively.
When we choose the mass solution for $g=0$ to 
coincide with the corresponding result in the 3+1-dimensional theory the behavior of the results of the renormalized and unrenormalized systems coincide.

\begin{figure}
\centering
\null\hfill
\includegraphics[width=0.45\textwidth]
{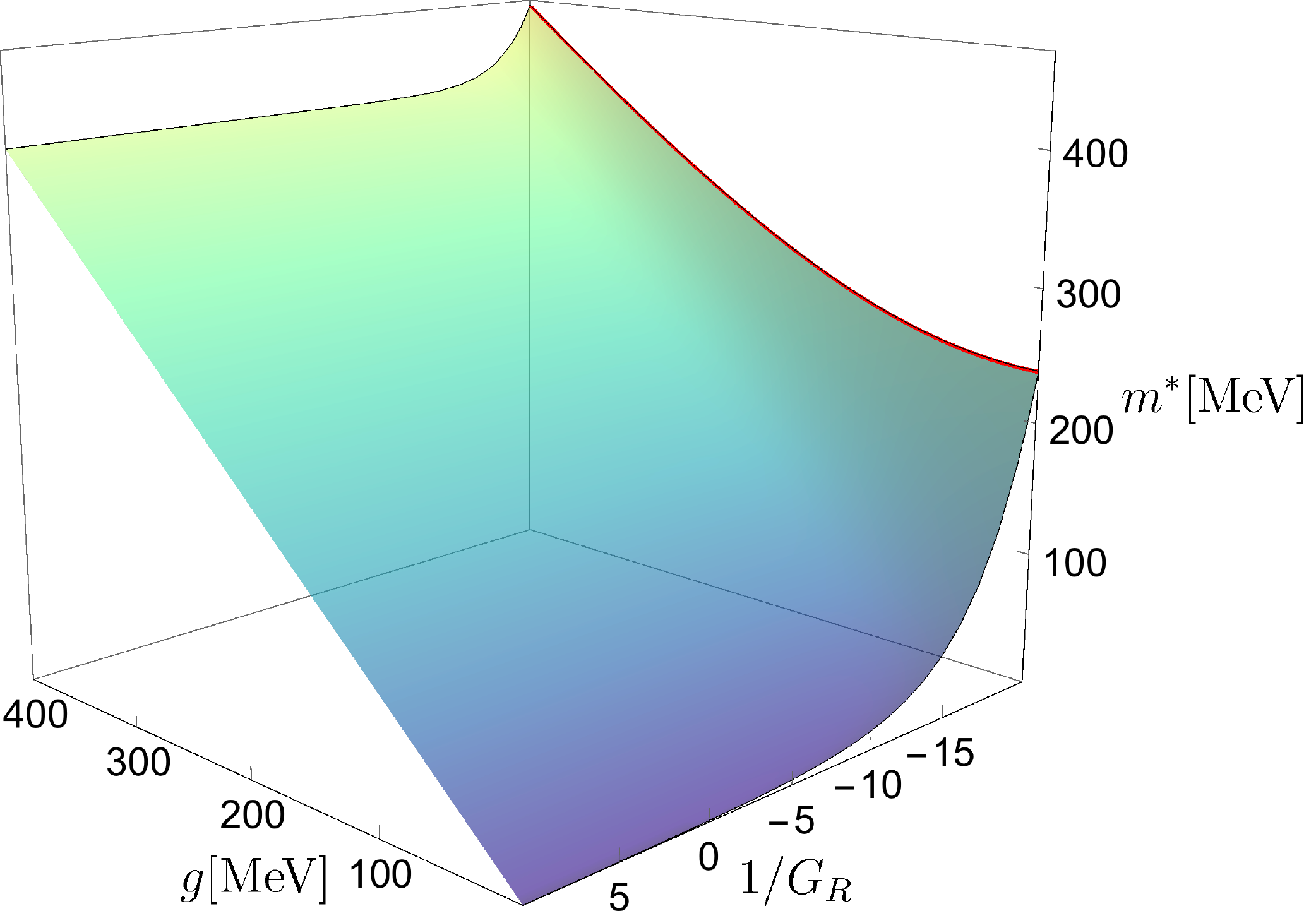}
\hfill\null
\caption{
\label{f6} 
Behavior of the solution $m^*$ of the renormalized gap equation (\ref{s3e21})
as a function of the coupling constant $g$ and the renormalized four-point   
interaction strength $G_R$. The behavior of the solution that coincides with 
the result of the  3+1-dimensional theory for $g=0$ is shown in red.
}
\end{figure}

Figure \ref{f6} shows the general behavior of the solution (\ref{s3e22}) to the 
renormalized gap equation as a function of the coupling constant $g$ and 
the renormalized four-point interaction strength $G_R$.
The behavior of the solution that coincides with the result of the 
3+1-dimensional theory for $g=0$ is shown in red and can be compared 
with the solution to the unrenormalized gap equation shown in Fig.~\ref{f4}. 

\section{Concluding remarks}
\label{s4}

$\cPT$ symmetry is understood as the complex extension of Hermitian quantum theory 
\cite{BBJ}. 
Here, we have investigated the non-Hermitian $\cPT$-symmetric extension of the 
Nambu--Jona-Lasinio model in 3+1 and 1+1 dimensions, and studied the effects of 
these non-Hermitian terms on the process of mass generation.  
Our major results are the following: 
(1) In previous calculations for the Dirac equation that include 
non-Hermitian bilinear terms, contrary to expectations, no real mass spectra can be 
obtained in the chiral limit; a nonzero bare fermion mass is essential for the 
realization of $\cPT$ symmetry in the unbroken regime. 
Here, in the NJL model, in which four-point interactions are present, we 
{\it do} find real values for the mass spectrum also in the limit of vanishing 
bare masses in both 3+1 and 1+1 dimensions, at least for certain specific 
values of the non-Hermitian couplings $g$. 
Thus, the four-point interaction overrides the effects leading to $\cPT$ 
symmetry-breaking for these parameter values. 
(2) In 3+1 dimensions, we note that we are able to introduce a non-Hermitian 
bilinear term that {\it preserves} the chiral symmetry of the model; in 1+1 
dimensions, however, the only non-Hermitian $\cPT$-symmetric term does not possess chiral 
symmetry. 
However, in both cases, the non-Hermitian term leads to a change in the 
generated mass. 
In both models, this can be tuned to be small; we can fix the bare fermion mass to 
its  value when $m_0=0$ in the absence of the non-Hermitian term, and thus
determine the small generated bare fermion mass. 
(3) In both cases, the gap equations display a rich phase structure as a 
function of the coupling strengths.

\end{document}